\documentclass[floatfix,twocolumn,showpacs,preprintnumbers,amsmath,amssymb,superscriptaddress]{revtex4}
\usepackage{graphicx,psfrag,subfigure}
\usepackage{amsfonts}

\def\simlt{\lower.5ex\hbox{$\; \buildrel < \over \sim \;$}}
\def\simgt{\lower.5ex\hbox{$\; \buildrel > \over \sim \;$}}
\def\simpropto{\lower.2ex\hbox{$\; \buildrel \propto \over \sim \;$}}

\newcommand{\newc}{\newcommand}
\newc{\gsim}{\lower.7ex\hbox{$\;\stackrel{\textstyle>}{\sim}\;$}}
\newc{\lsim}{\lower.7ex\hbox{$\;\stackrel{\textstyle<}{\sim}\;$}}
\newc{\gev}{\,{\rm GeV}}
\newc{\mev}{\,{\rm MeV}}
\newc{\ev}{\,{\rm eV}}
\newc{\kev}{\,{\rm keV}}
\newc{\tev}{\,{\rm TeV}}

\newc{\mz}{M_Z}
\newc{\mpl}{M_*}
\newc{\mw}{m_{\rm weak}}
\newc{\nr}[1]{N^c_R{}_{#1}}
\usepackage{amsmath}
\def\beq{\begin{equation}}
\def\eeq{\end{equation}}
\def\bea{\begin{eqnarray}}
\def\eea{\end{eqnarray}}
\def\bitem{\begin{itemize}}
\def\eitem{\end{itemize}}
\newc{\ie}{{\it i.e.}}          \newc{\etal}{{\it et al.}}
\newc{\eg}{{\it e.g.}}          \newc{\etc}{{\it etc.}}
\newc{\cf}{{\it c.f.}}
\def\inv{^{\raise.15ex\hbox{${\scriptscriptstyle -}$}\kern-.05em 1}}
\def\lbar{{\lower.35ex\hbox{$\mathchar'26$}\mkern-10mu\lambda}} 

\def\to{\rightarrow}

\let\<=\langle
\let\>=\rangle

\let\+=\uparrow

\begin{document}
\thispagestyle{empty}

\begin{center}
\title{Numerical estimation of the escaping flux of massless particles created in collisions around a Kerr black hole}
\date{May 10, 2011}
\author{Andrew J. Williams}
\email{williams@pheno.pp.rhul.ac.uk}
\affiliation{Royal Holloway, University of London, Egham, TW20 0EX, UK}
\affiliation{Rutherford Appleton Laboratory, Chilton, Didcot, OX11 0QX, UK} 
\begin{abstract}
The geodesics of massless particles produced in collisions near a rotating black hole are solved numerically and a Monte Carlo integration of the momentum distribution of the massless particles is performed to calculate the fraction that escape the black hole to infinity. A distribution of in falling dark matter particles, which are assumed to annihilate to massless particles, is considered and an estimate of the emergent flux from the collisions is made. The energy spectrum of the emergent particles is found to contain two Lorentz shifted peaks centred on the mass of the dark matter. The separation of the peaks is found to depend on the density profile of the dark matter and could provide information about the size of the annihilation plateau around a black hole and the mass of the dark matter particle.
 \end{abstract}
\maketitle
\end{center}

\section{Introduction}
\label{sec:intro}

Intermediate mass black holes may be associated with a density spike of dark matter \cite{BFTZ2009} that can provide an enhancement in the annihilation rate of dark matter particles to gamma rays or neutrinos. The indirect detection of dark matter through its annihilations could provide information about the nature of the dark matter in the galaxy and its distribution. The physics of particle collisions in the gravitational field of a black hole has been extensively studied, see e.g. \cite{Baushev} \cite{BSM} \cite{Berti:2009bk} \cite{Jacobson:2009zg} \cite{GP2010} \cite{Wei:2010vca} \cite{Zaslavskii:2010jd} \cite{Harada:2010yv} \cite{Kimura:2010qy}. In the case of rotating black holes, collisions between dark matter particles can yield large centre of mass energies \cite{BSM} and inside the ergosphere particles produced via the Penrose process could carry high energies \cite{GP2010}. In order to make use of any such collisions around a black hole it will be important to understand the fate of the particles produced, in particular the fraction of particles produced that escape the black hole compared to those which cross the event horizon. The escape fraction can be defined as the fraction of particles that escape having been produced in a collision of two dark matter particles. The escape fraction depends on the momenta of the colliding dark matter particles and the distance from the black hole at which they collide. The escape fraction has been calculated for maximally rotating black holes where massless particles are emitted in the equatorial plane of the black hole \cite{escapefrac} and is also known for Schwarzschild black holes \cite{mtw}. However a general analytical solution is not known for rotating black holes where the particles are not restricted to the equatorial plane. 

The goal of this paper is to find a numerical result for the escape fraction of massless particles produced in collisions around a rotating black hole and to investigate the energy spectrum of any emergent flux. We assume an interaction of the form $\chi \chi \rightarrow x x$ where $x$ is a massless particle that could be a photon or a massless neutrino. (The analysis can also be used for neutrinos with small masses as the change in the result is negligible provided the mass of the dark matter is much larger than the neutrino mass.) To simplify the analysis, we assume that the final state particles are emitted isotropically in the centre of mass frame of the collision.

The paper is organised as follows: in Section~\ref{sec:geodesic} the method of numerically solving the geodesic equation for a massless particle in a Kerr metric with arbitrary rotation is presented. Section~\ref{sec:escapefrac} uses this to obtain the escape fraction for massless particles and introduces the effect of a boosted frame of reference due to the momentum of the colliding particles. In Section~\ref{sec:produced} the distribution of the momenta of the colliding dark matter particles is outlined. Section~\ref{sec:spectrum} presents the spectrum of emergent particles from the collisions and finds that the spectrum generally contains two Lorentz shifted peaks centred around the dark matter mass.  In Section~\ref{sec:conclusion} we present our conclusions.
\section{Numerical solution of the geodesic equation}
\label{sec:geodesic}
To know whether a particular particle escapes the black hole or not, its geodesic equation must be solved and the behaviour as $t \to \infty$ found. This is most easily performed using Boyer-Lindquist coordinates as defined in the analysis of Ref.~\cite{BPT}, which is followed closely here. In these coordinates the metric for a Kerr black hole has the form \cite{BPT}, 
\begin{multline}
 ds^2 = -\left(1 - \frac{2Mr}{\Sigma}\right)dt^2 - \left(\frac{4Mar \sin^2\theta}{\Sigma}\right) dtd\phi \\
 + \frac{\Sigma}{\Delta}dr^2 + \Sigma d\theta^2 + \left(r^2 + a^2 + \frac{2Ma^2r \sin^2\theta}{\Sigma}\right) \sin^2\theta d\phi^2,
\end{multline}
where $M$ is the mass of the black hole, $a$ is the angular momentum per unit mass of the black hole, $\theta$ is the polar angle from the axis of rotation, $\phi$ is the azimuthal angle and the following functions are defined \cite{BPT},
\bea
\Delta &\equiv& r^2 - 2 M r + a^2, \\
\Sigma &\equiv& r^2 + a^2\cos^2\theta.
\eea
For simplicity the mass of the black hole is set to $M = 1$, as a result $a$ ranges between $0$ and $1$ in units where $\frac{M_{BH} G}{c^2} = 1$ such that the Schwarzchild radius $r_s = 2$. The trajectories of a particle are then defined by 3 constants of motion which are conserved, the forms of these quantities are \cite{BPT}
\begin{eqnarray}
  \nonumber
 E &=& -p_t,\,\,\,\,\,\,\,\,\,\,\,\,\,\,\,\,\,\,\,\,\,\,L = p_\phi, \\
 Q &=& p_\theta^2 + \cos^2\theta\left(a^2\left(\mu^2 - p_t^2\right) + \frac{p_\phi^2}{\sin^2\theta}\right). 
\end{eqnarray}
Here $\mu$ is the mass of the particle, $E$ is the energy, $L$ is the component of the angular momentum parallel to symmetry axis of the black hole and $Q + p_\phi^2$ is the total angular momentum squared when $a = 0$. $Q$ characterises the motion in the $\theta$ direction and for $Q = 0$ particles in the equatorial plane will remain restricted to that plane \cite{BPT}. The equations of motion for the particles can now be written in terms of $\lambda$, which will become an affine parameter for massless particles, and are written here in the form given by Ref.~\cite{BPT},
\begin{eqnarray}
  \nonumber
  \Sigma \frac{dr}{d\lambda} &=& \pm V_r = \pm \sqrt{T^2 - \Delta\left(\mu^2r^2 + \left(L - aE\right)^2 + Q\right)}, \\
  \nonumber
  \Sigma \frac{d\theta}{d\lambda} &=& \pm V_\theta = \pm \sqrt{Q - \cos^2\theta\left(a^2\left(\mu^2 - E^2\right) + \frac{L^2}{\sin^2\theta}\right)},\\
  T &=& E\left(r^2 + a^2\right) - La.
\end{eqnarray}
The final state particles of the collisions are assumed to be massless allowing the equations to be simplified by setting $\mu = 0$ (For massive particles the approximation $\mu = 0$ is good provided $\mu \ll E$ and $E$ is generally of the order of the dark matter mass. The results therefore remain unchanged if a small mass is introduced.). In order to calculate the escape function numerically the equations of motion for $r$ and $\theta$ were integrated numerically until the particle either crossed the horizon or escaped to some large value of $r$. The algorithm employed was based on an embedded adaptive step size Runge-Kutta formula as outlined in Ref.~\cite{numerical}. 

The integration was ended whenever the massless particle crossed the horizon or achieved some maximum distance from the black hole, for example $r_{max}\approx 100$ for collisions close to the horizon. This was increased for collisions further from the black hole. In the case of bound orbits the integration was ended after a suitable number of turning points in the $r$ solution or a maximum value for $\lambda$ was reached. In practise the parameters of the algorithm, the minimum step size, $\lambda_{max}$, $r_{max}$ and the value of $\frac{dr}{d\lambda}$ that indicates a turning point were varied to find optimal values that did not introduce significant errors in the final result. This was checked against test cases and known solutions for the escape fraction\cite{mtw}\cite{escapefrac} shown in Figures.~\ref{fig:a=0} and \ref{fig:a=1}.

\section{The escape fraction}
\label{sec:escapefrac}
This method for solving the geodesic equation for the emergent particle then allows the escape fraction of massless particles produced in collisions around the black hole to be found. The escape fraction was calculated by a Monte Carlo integration over the final momentum of the produced massless particle pair. In the centre of mass frame of the colliding particles the massless particles are emitted back to back and carry energy equal to one half of the centre of mass collision energy. The momentum integration is therefore reduced to an integration over the direction of the massless particle, which we define in terms of two angles, $\theta$ the angle from the radial axis parallel to the polar angle of the black hole coordinate system and $\phi$ the azimuthal angle about the radial axis. 

For the integration only one of the pair of massless particles was considered and a flat distribution of its angles in the centre of mass frame was generated. For each generated direction the geodesic equation was solved numerically as above and the massless particle was recorded as either escaping or becoming trapped by the black hole. The escape fraction was then taken as the number of massless particles recorded as escaping divided by the total number generated.

The uncertainty in the escape fraction due to the statistical effects of the Monte Carlo integration was found by separating the massless particles into discrete bins, the escape fraction for each bin was determined then the average escape fraction taken. The variance in the escape fraction for each bin then gave an estimate for the uncertainty in the averaged escape fraction. This uncertainty does not account for any error in the numerical solution for the geodesic equations but since the escape fraction only depends on the general behaviour of each massless particle this was neglected.

To test the implementation of the numerical method the escape function can be calculated for the $a = 0$ Schwarzschild case. Analytically the conditions for escape are known, and can be converted to an escape fraction as a function of the radial position \cite{mtw}.
\begin{figure}[h!]
\centerline{\includegraphics[width=8cm]{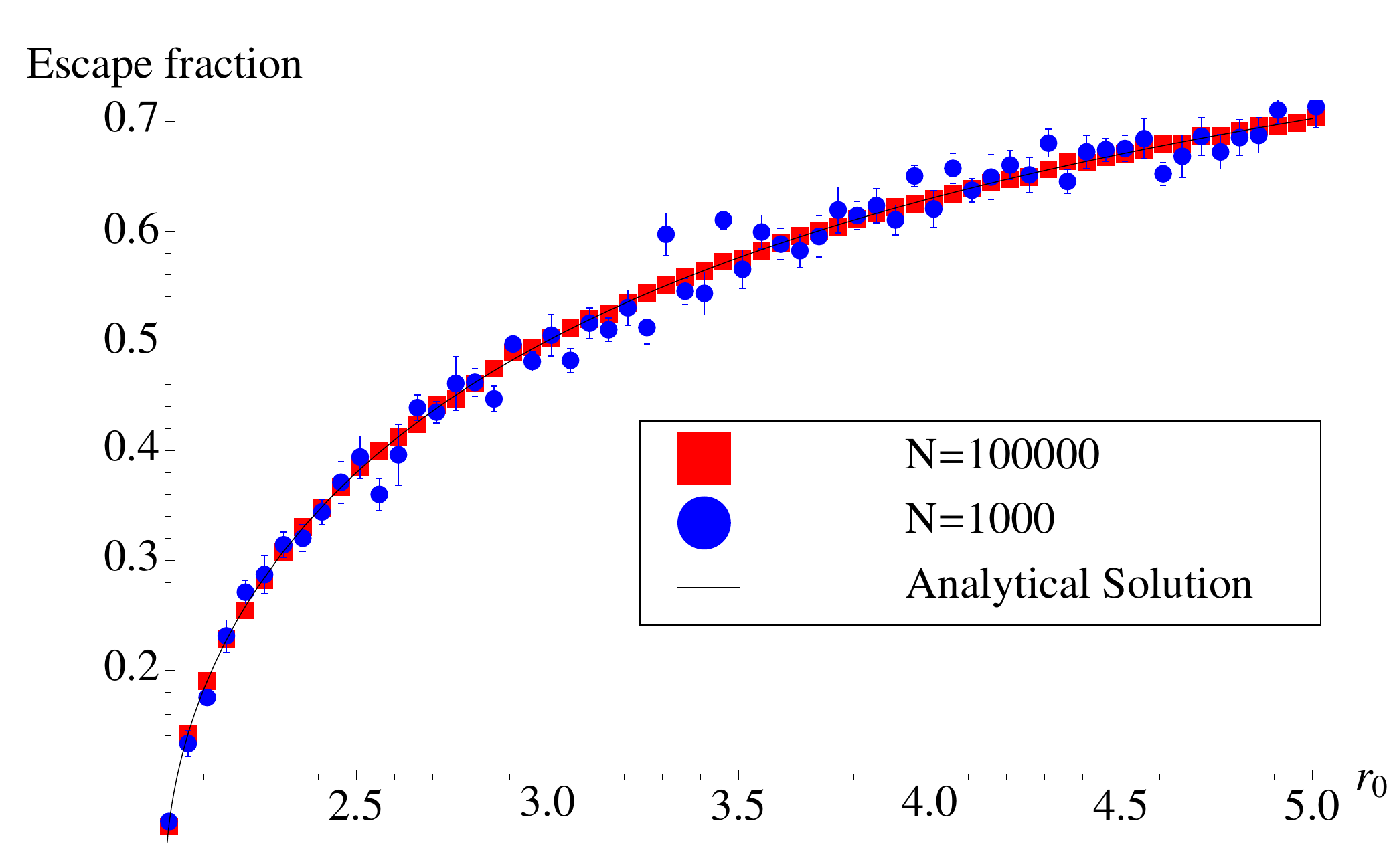}}
 \caption{The escape fraction for a = 0, where N is the number of Monte Carlo iterations. The solid line shows the escape fraction constructed from the analytical conditions of escape}
\vspace{-0.15cm}
 \label{fig:a=0}
\end{figure}
Fig.~\ref{fig:a=0} shows the behaviour of the escape fraction when the massless particle is emitted in an unboosted frame so that the emission is isotropic in the lab frame. The escape fraction matches the function determined from \cite{mtw} with some variation due to the Monte Carlo integration. The effect of the number of Monte Carlo iterations on the error in the escape fraction can be seen and gives an indication of the number of iterations required to give a good agreement with the analytical solution. 
\subsection{Locally non-rotating and boosted frames}
Having tested the machinery for the non-rotating case results can be obtained for black holes with $a > 0$. For $a = 1$ it will  be useful to consider two cases for the numerical calculation, firstly where the massless particles are generated with no component of their motion outside the equatorial plane ($Q = 0$) and secondly where they are generated in all directions to compare with the result found in \cite{escapefrac}. For a rotating black hole there is another complication. So far the massless particles have been assumed to be emitted in a locally flat frame that is at rest in the Boyer-Lindquist coordinate system. For a rotating black hole the corresponding frame is that of a locally non rotating frame (LNRF) \cite{BPT}. A LNRF represents a set of Minkowski coordinates such that the physics in this local frame can be described by special relativity. This is further complicated when the centre of mass frame of the collision is Lorentz boosted relative to the LNRF. In this case it is easiest to generate the massless particle momentum in the centre of mass frame where it is assumed to be isotropic and apply a Lorentz transformation to attain the momentum in the LNRF. 

The transformation from the isotropic distribution of momentum in the centre of mass frame to the distribution of the constants of motion in the Boyer-Lindquist coordinate system depends on the coordinates at which the collision takes place, $r_0$ and $\theta_0$ and $\Lambda(P_1, P_2)$ the Lorentz transformation between the centre of mass frame and the LNRF. Where $P_1$ and $P_2$ are defined as the momenta of the colliding particles in the LNRF. For details on transforming between a LNRF and a Boyer-Lindquist system see Ref.~\cite{BPT}, while the method of calculating $\Lambda(P_1, P_2)$ we follow that in Ref.~\cite{escapefrac}.

The general procedure for the calculation of the massless particle momentum in the Boyer-Lindquist coordinates was carried out as follows. Firstly the momenta of the incoming particles were defined in the Boyer-Lindquist frame by setting the constants of motion $L$ and $Q$ then evaluating the equations for their momentum at the coordinates that the collision was assumed to take place. The incoming momenta were transformed to the LNRF as in Ref.~\cite{BPT}. With the two momenta of the incoming particles defined in this locally flat space the relative momentum of the centre of mass frame was found and the boost between these two frames determined. The massless particle direction was then randomly generated in the centre of mass frame, the boost between the LNRF and the centre of mass frame was inverted to allow the massless particle momentum in the LNRF to be found. Finally the massless particle momentum in the Boyer-Lindquist coordinates was calculated giving the constants of motion for the massless particle. The geodesic equation for the massless particle could then be integrated numerically as before. 

 The momentum of the incoming particles was determined following the procedure in Ref.~\cite{escapefrac} which calculated the velocities of the incoming particles that were restricted to the equatorial plane. The momenta of the two incoming particles was then used to produce the 4 dimensional Lorentz boost between the LNRF and centre of mass frame.
\begin{figure}[h!]
\centerline{\includegraphics[width=8cm]{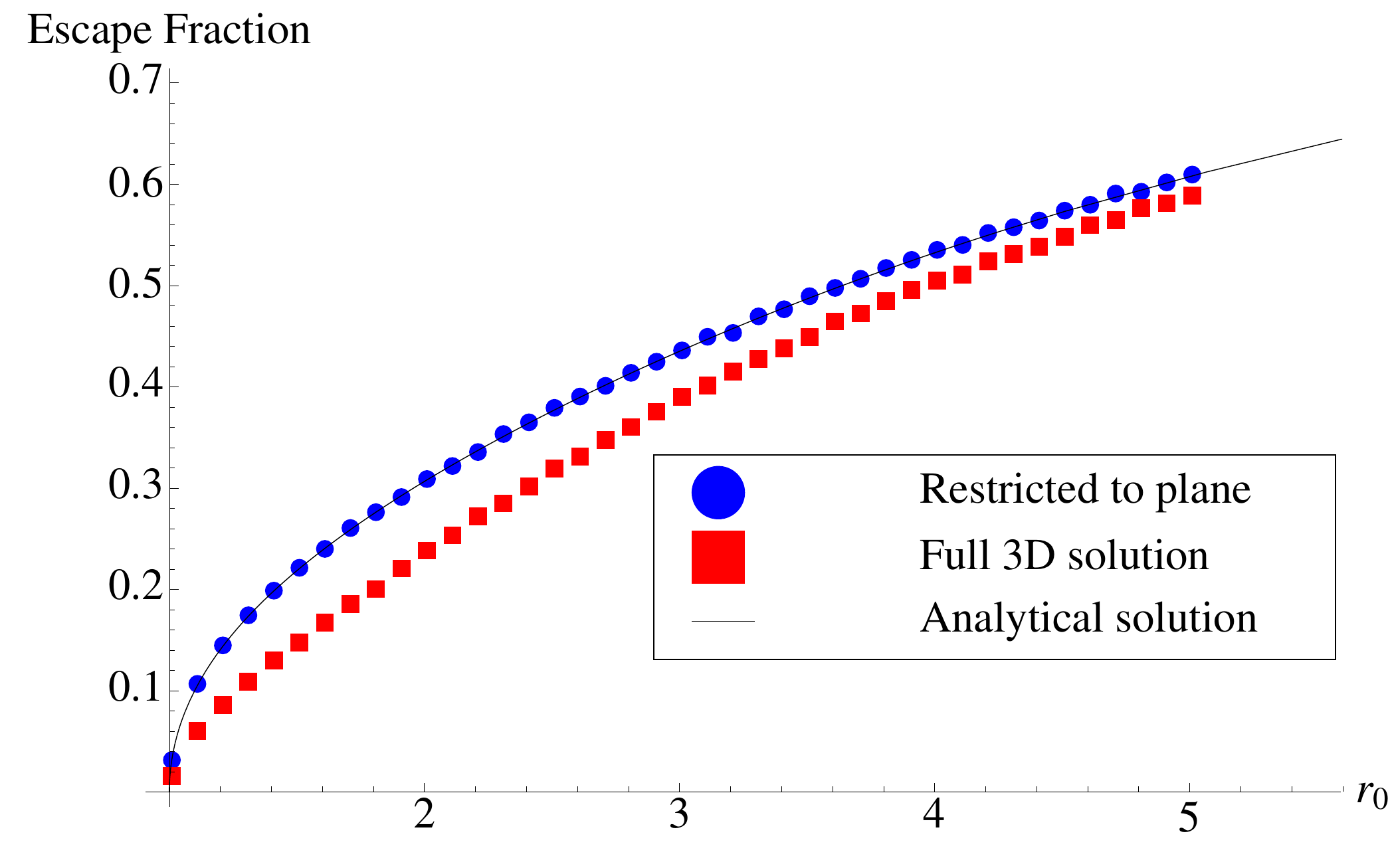}}
 \caption{Escape fraction for a=1 with momenta of colliding particles fixed: Analytic solution in equatorial plane (Black line) \cite{escapefrac}, Numerical result in equatorial plane (blue circles) and full numerical result allowing propagation in all directions (red squares).}
\vspace{-0.15cm}
 \label{fig:a=1}
\end{figure}
Fig.~\ref{fig:a=1} shows the escape fraction as a function of $r$ for collisions between two massive particles colliding in the equatorial plane of a black hole with $a = 1$. The colliding particles have angular momentum $L_1 = 2$ and $L_2 = -2$ which correspond to the constant of motion $L$ and both have $Q = 0$. The escape fraction found through an analytical calculation \cite{escapefrac} when restricted to the equatorial plane is also shown. The numerical result matches the exact solution when the particles are confined to the equatorial plane. The numerical result when the massless particles are emitted in all directions gives a value for the escape fraction which is comparable in size to the result in equatorial plane, suggesting that the approximation used in Ref.~\cite{escapefrac} is reasonable.
\section{Distribution of momenta of the colliding particles}
\label{sec:produced}
The previous section considered massless particles produced in collisions of massive particles with fixed incoming momenta however the escape fraction depends on the Lorentz boost from the LNRF to the centre of mass frame given above, which is derived from the constants of motion $E$, $L$ and $Q$ of the incoming particles. These constants are not fixed in general and the momenta of the colliding particles will have some distribution. This distribution will also affect the centre of mass energy of the collision and the energy of the produced massless particles as well as the direction in which they are emitted. In order to calculate the energy distribution of escaping particles and the general escape fraction it will therefore be important to consider incoming particles with some distribution of momenta.

\begin{figure}[h!]
\vspace{0cm}
\centerline{\includegraphics[width=8cm]{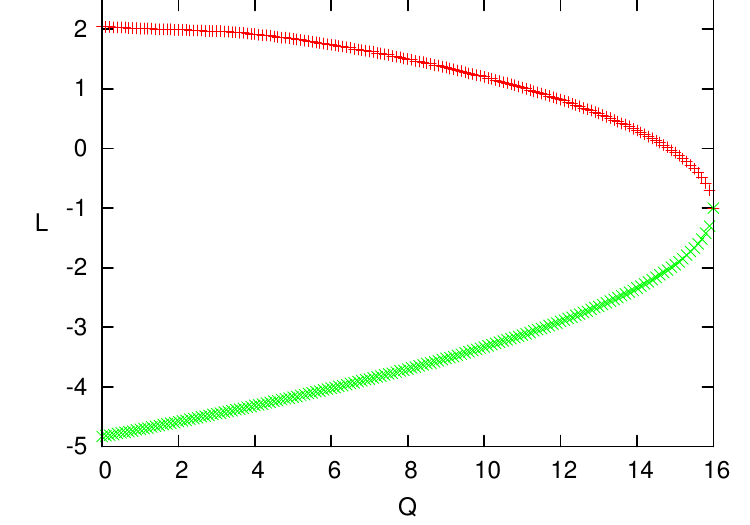}}
\vspace{0.5cm}
\caption{Allowed range in $Q$ and $L$ for incoming particles that reach $r_0=1.1$ with $a = 1$. The red line show $L_{max}$ and the green line $L_{min}$. \label{fig:inrange}}
\end{figure}

 The form of this distribution is not known so as a first approximation we assume a flat distribution in the constants L and Q. The allowed range of L and Q was fixed by considering only particles with geodesics that would allow the particle to fall from some large r to the radius $r_0$ at which the collision takes place. The constant of motion $E$ was set to the mass of the dark matter which in turn is set to 1. The range of allowed constants of motion was calculated by fixing $\theta$ to the equatorial plane then calculating $\frac{dr}{d\lambda}$ as a function of r starting at $r\gg r_0$ then running down in small steps of r to the value $r_0$, if $\frac{dr}{d\lambda}$ becomes very small then we assume that a turning point is reached and that a particle could not fall from infinity to $r_0$. The distribution of allowed momenta was sampled by Monte Carlo integration. Figure.~\ref{fig:inrange} shows the allowed range in $L$ and $Q$ for collisions occurring at $r_0 = 1.1$ as an example. This is characterised (for $a = 1$) by a maximum value of $Q$ which occurs for $L=-1$ as well as a maximum and minimum value of $L$ with the largest range corresponding to $Q=0$.
\begin{figure}[h!]
\vspace{0cm}
\centerline{\includegraphics[width=8cm]{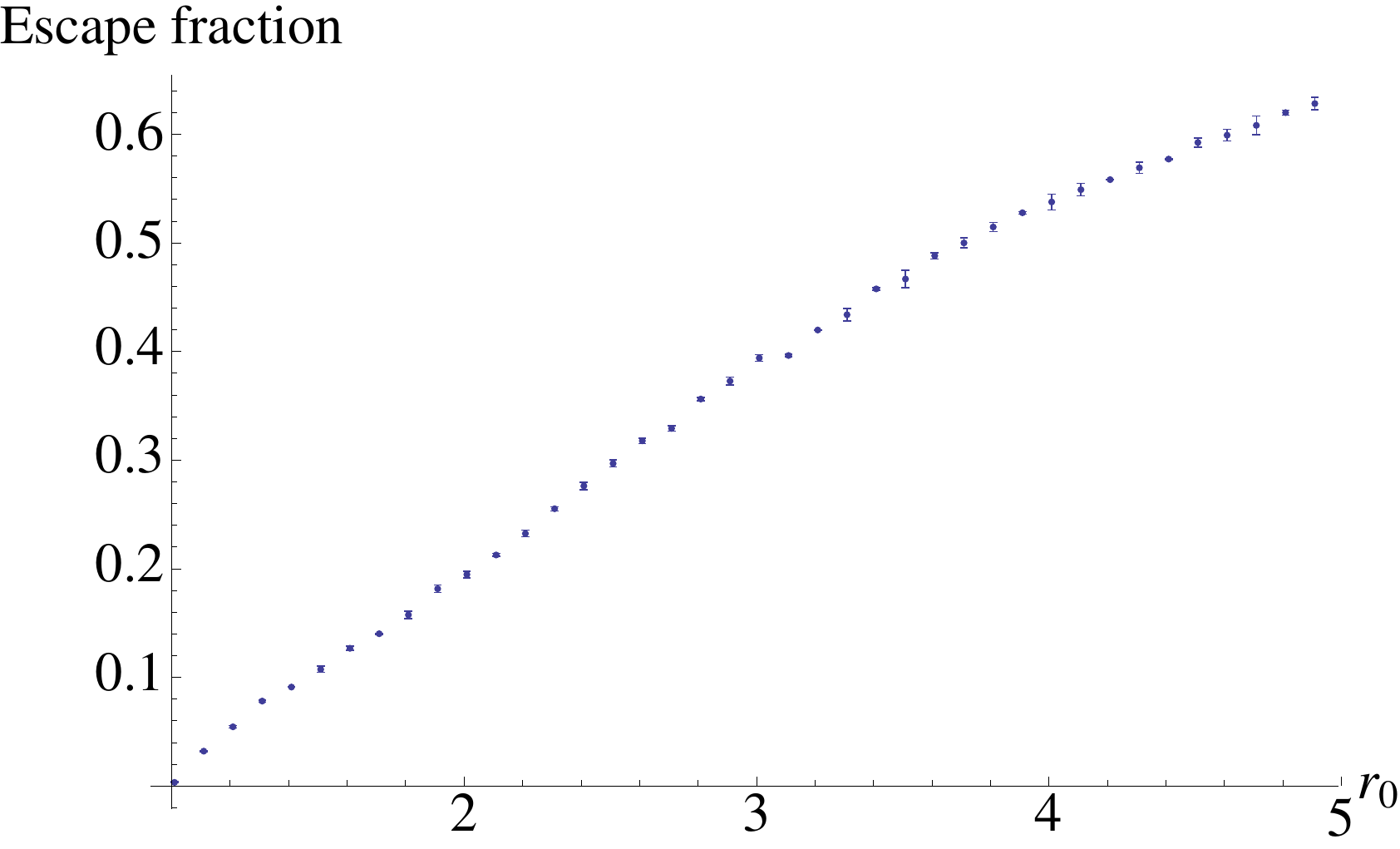}}
\caption{Escape fraction for $a = 1$ with integration over momenta of the colliding particles including the effect of the Lorentz boost.\label{fig:escapefull}}
\end{figure}
Figure.~\ref{fig:escapefull} shows the escape fraction when the momenta of the colliding particles is averaged over all allowed values of $L$ and $Q$. The escape fraction has only a weak dependence on the Lorentz boost and hence L and Q, of the colliding particles so the full escape fraction is only a slight shift from the result found for a particular fixed boost shown in Figure.~\ref{fig:a=1}. 
\section{Spectrum of emergent massless particles}
\label{sec:spectrum}
In order to calculate the energy spectrum of emergent particles we must integrate the effect of collisions over the distribution of dark matter surrounding the black hole. The flux arriving at some distance $D$, from the black hole reads \cite{escapefrac},
\begin{equation}
 \Phi\approx\frac{\sigma v r_s^3}{4 \pi m_\chi^2 D^2}\, \int_{r_h}^{r_\infty}\,  \rho^2(r) \;e(r)\, dV, 
\end{equation}
where $\sigma v$ is the cross-section for annihilation to massless particles which we assume to be energy independent, $m_\chi$ is the mass of the dark matter particles, $r_s$ is the Schwarzschild radius, $r_h$ is the event horizon, $e(r)$ is the escape fraction shown in Figure.~\ref{fig:escapefull} and $\rho(r)$ is the dark matter density around the black hole. It is assumed that the escape fraction is spherically symmetric. The calculation is performed for collisions occurring in the equatorial plane only, to simplify the integration. By writing 
\begin{equation}
\label{eq:rho}
 \rho(r) = \rho_{pl} \rho_0(r),
\end{equation}
 where $\rho_0(r)$ is a dimensionless function giving the shape of the density profile. The flux can be written in terms of a dimensionless integral containing all the of the dependence on the rotation of the black hole and the shape of the density profile as \cite{escapefrac},
\begin{eqnarray}
 \Phi&\approx&\frac{\sigma v r_s^3 \rho_{pl}^2}{m_\chi^2 D^2}\, {\cal{I}}(a, r_1), \\ \label{eq:PHI}
 {\cal{I}}(a, r_1)&=& \int_{r_h}^{r_\infty}\, r^2\, \rho_0^2(r) \;e(r)\, dr.
\end{eqnarray}
In order to calculate the spectrum of the emergent massless particles we need to know $\Phi(E)$ which depends on the convolution of the escape fraction and spectrum of massless particles produced in the collisions. To find this we define $p(r, E_1, E_2)$ as the fraction of massless particles that escape with energy between $E_1$ and $E_2$ per collision. This was found numerically by separating the escaped massless particles into bins of energy and dividing by the total number of collisions. The integral can then be rewritten as 
\begin{equation}
 \Phi(E_1, E_2) \propto {\cal{I}}(a, r_1, E_1, E_2) = \int_{r_h}^{r_\infty}\, r^2\, \rho_0^2(r) \;p(r, E_1, E_2)\, dr,
\end{equation}
where summing over all the energy bins gives the total flux.

The density distribution of dark matter around black holes has been extensively studied for non-rotating black holes see \cite{BFTZ2009}  \cite{Bertone:2005xv} \cite{Bertone:2005xz}. Considering the case of an intermediate mass black hole the density distribution can be described as follows \cite{BFTZ2009}; close to the black hole there is an annihilation plateau with constant density $\rho_{pl} = m_\chi/(\sigma v\, t)$ where $t$ is the formation time of the black hole. This holds out to a radius $r_{cut}$ from which point the density falls off with a power law $\rho \propto r^{-\frac{7}{3}}$. The distribution here is assumed to be spherically symmetric (For a rotating black hole this may no longer be true but serves as a first approximation.). The flux will be dominated by contributions from the annihilation plateau and density spike since the escape fraction is only small for radii much smaller than the annihilation plateau. It is now clear that if $\rho_{pl}$ given in Eq.~\ref{eq:rho} is taken to be the density of the annihilation plateau, $\rho_0(r)$ will take the form \cite{BFTZ2009},
\beq\rho_0(r) = \left\{\begin{array}{l l}
  1 & \quad \mbox{if $r < r_{cut}$}\\
  \left(\frac{r}{r_{cut}}\right)^{-\frac{7}{3}} & \quad \mbox{if $r > r_{cut}$.}\\ \end{array} \right. \eeq

To calculate the value of ${\cal{I}}$ for a particular set of black hole parameters the value of $r_{cut}$ needs to be specified. In general this depends on the mass of the black hole and on the annihilation cross-section of the dark matter. For the super massive black hole at the centre of the galaxy this was estimated to be $r_{cut} \approx 4\times10^{-5}pc$ from Ref.~\cite{Bertone:2005xv} which is $r_{cut} \approx 137r_s$ in terms of the Schwarzschild radius. 

\begin{figure}[h!]
\vspace{0cm}
\centerline{
\subfigure[ $r_{cut} = 5$]{
\includegraphics[width=8cm]{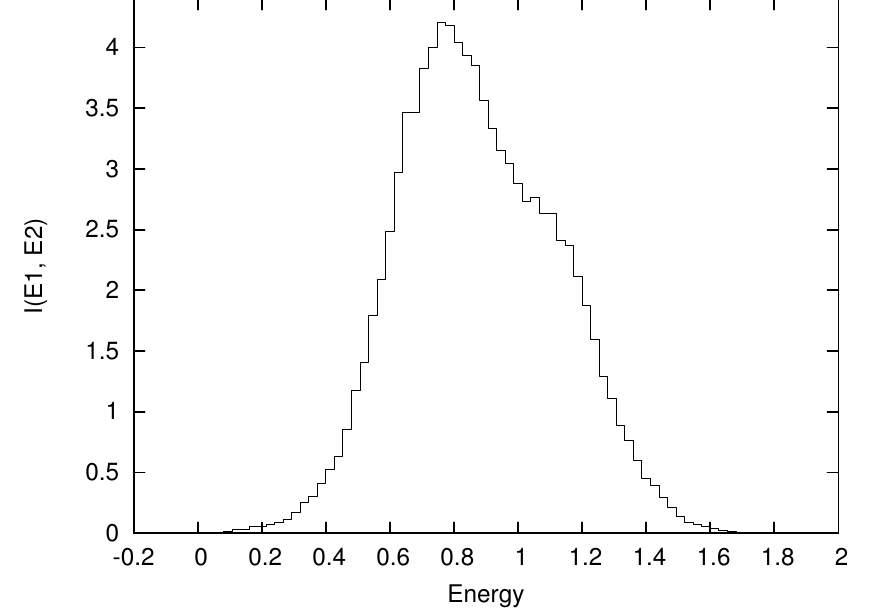}
\label{fig:inspec1-5}
}
}
\vspace{0.1cm}
\centerline{
\subfigure[ $r_{cut} = 34$]{
\includegraphics[width=8cm]{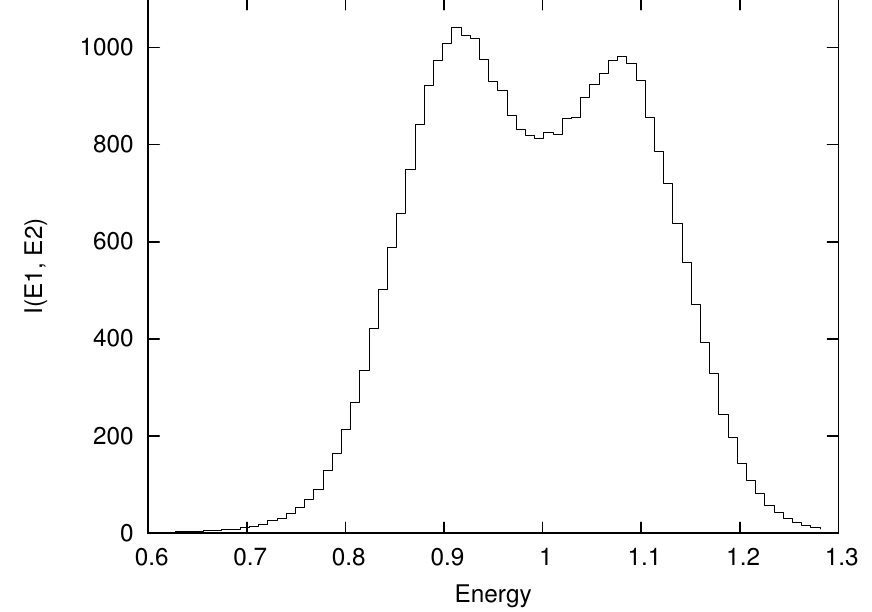}
  \label{fig:inspec1-34}
}
}
\vspace{0.2cm}
\centerline{
\subfigure[ $r_{cut} = 274$]{
\includegraphics[width=8cm]{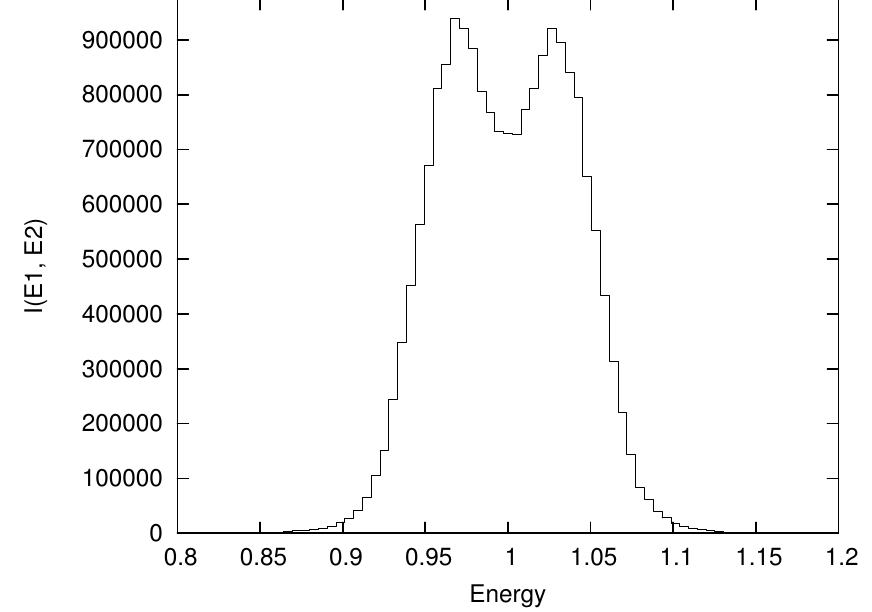}
  \label{fig:inspec1-274}
}
}
\caption{Spectrum of massless particles escaping to infinity with a cut off in dark matter density at various values of $r_{cut}$. ($a=1$) \label{fig:escaped}}
\label{fig:inspec}
\end{figure}

The exact value is not important since the shape of the spectrum remains distinctive over a large range of $r_{cut}$ and a measurement of the spectrum should in principle allow $r_{cut}$ to be determined. To show this, the spectrum produced for a number of different values of $r_{cut}$ are shown in Figure.~\ref{fig:inspec}. The energy is expressed in units of the dark matter mass.

Figure.~\ref{fig:inspec1-5} shows the spectrum for a small value of $r_{cut} = 5$ (in units where $r_s = 2$), and hence is dominated by collisions close to $r_0 = 5$. Here the spectrum is characterised by a single asymmetric peak below the dark matter mass. The peak itself is broad and the mass of the dark matter is found to be roughly equidistant from the peak and the edge of the shoulder in the distribution. This structure arises due to several effects. Firstly there is a balance between the centre of mass energy and the gravitational red shift as suggested in Ref.~\cite{Baushev}, this is expected from energy conservation considerations that the energy available to the collision as measured far from the black hole will be $2 m_\chi$. 

The result is that the spectrum is not shifted to higher energies by the increasing centre of mass energy close to the black hole horizon or red shifted by increasing gravitational potential. There is a shift of the peak to lower energies that arises from the Lorentz boost, since the centre of mass frame is in general boosted towards the horizon, massless particles emitted away from the black hole are red shifted to lower energies. Particles emitted towards the horizon are correspondingly shifted to higher energies. The effect of this is to split the spectrum of escaping particles into two peaks, a red shifted peak with energy below the dark matter mass and a blue shifted peak with energy above the dark matter mass.

In Figure.~\ref{fig:inspec1-5} only one such peak is visible this is because particles emitted towards the horizon are much less likely to escape the black hole. There is still a slight shoulder to the spectrum just above the dark matter mass which corresponds to the remnant of the blue shifted peak after the escape fraction has been taken into account. 

As $r_{cut}$ in Figure.~\ref{fig:inspec1-34} increases to $34$, both peaks become clearly visible and the spectrum becomes less broad. The height difference between the peaks also decreases as they get closer together. The peaks in the spectrum move towards the dark matter mass since the Lorentz boost to the centre of mass frame becomes smaller. The size of the Lorentz boost decreases as the average radial momenta of the colliding particles decreases. The escape fraction becomes less important as $r$ increases and the massless particles initially emitted towards the black hole can now escape in many cases. It is still more likely that a massless particles emitted away from the horizon will escape than one emitted towards it so the blue shifted peak is generally smaller than the red shifted one. It also becomes clear that the spectrum is indeed roughly symmetric around the dark matter mass which could be useful in determining the mass from the spectrum. 

In Figure.~\ref{fig:inspec1-274} the spectrum is shown for $r_{cut} = 274 = 137r_s$  the collisions responsible for the majority of the flux are now occurring far from the horizon. The escape fraction becomes even less important and nearly all of the massless particles produced in the collisions will now escape. This can be seen in the spectrum by noting that the red shifted and blue shifted peaks are now very nearly the same height. Comparing the spectrum with that in Figure.~\ref{fig:inspec1-34} it can be seen that the red shifted and blue shifted peaks continue to move towards the dark matter mass as the effect of the Lorentz boost diminishes. 

The spectrum could potentially be used to infer some of the quantities in the system. The dark matter mass is indicated by the minimum between the peaks and the value of $r_{cut}$ can be estimated from the separation of the peaks as a fraction of the dark matter mass. 

The value of $r_{cut}$ can be estimated as follows; the dark matter mass must be found first by locating the minimum of the spectrum, the energy at which the first peak occurs as a fraction of the mass of the dark matter then decreases as a function of $r_{cut}$. By plotting the peak location as a function of $r_{cut}$ from simulation the expected value of $r_{cut}$ can be found for a given peak separation. It should be noted that the distribution is not exactly symmetric and that the higher energy peak is located closer to the dark matter mass than the lower energy peak. However both could be used to find an estimate of $r_{cut}$ and the ratio of the height of the peaks could also be utilised in this way. One problem with this analysis is that it assumes that the angular momentum of the black hole is known and varying $a$ shifts the separation of the peaks. However this effect is much smaller than changing $r_{cut}$ but would still introduce uncertainty in the estimation. The ability to resolve the peaks depends on the total flux available and the energy resolution around the dark matter mass of a particular measurement. In terms of the annihilation plateau the separation of the peaks was found up to $r_{cut} = 10^5$ with a separation of $\Delta_E \sim 0.001m_\chi$. For $r_{cut} = 10^3$ a separation of $\Delta_E \sim 0.03m_\chi$ was found. In the case that the peaks can not be resolved then the spectrum can be approximated as a single peak with a width comparable to the peak separation centred at the dark matter mass.

The total flux can be estimated by choosing some suitable values for the parameters in Eq.~\ref{eq:PHI}. Setting the mass of the black hole to $M=40\times10^5 M_\odot$, the annihilation cross-section of the dark matter as $\sigma v=10^{-28}\text{cm}^2\text{s}^{-1}$, the distance from the black hole $D=10\text{pc}$ and the time-scale for the growth of the black hole as $t_0=10^{10}\text{years}$ the flux can be written as \cite{escapefrac},
\begin{equation}
 \Phi = \Phi_0 {\cal{I}},
\end{equation}
where $\Phi_0 = 3.41 \text{km}^{-2} \text{year}^{-1}$. Integrating over $r$  for $r_{cut} = 274$ gives a total value of ${\cal{I}} \approx 2\times10^7$ which gives a total flux of $\Phi \approx 7\times10^7 \text{km}^{-2} \text{year}^{-1}$. For 10TeV dark matter annihilating to high energy gamma rays the HESS experiment could be sensitive to such a large flux however the energy resolution would be smaller than the separation of the peaks \cite{Aharonian:2006pe}. For smaller values of $r_{cut}$ resolution of the peaks becomes easier but the total flux is reduced. At $r_{cut} = 34$ the separation is potentially large enough to be observed but ${\cal{I}} \approx 10^4$ requiring greater sensitivity or observation time. 

\begin{figure}[h!]
\vspace{0.3cm}
\centerline{\includegraphics[width=8cm]{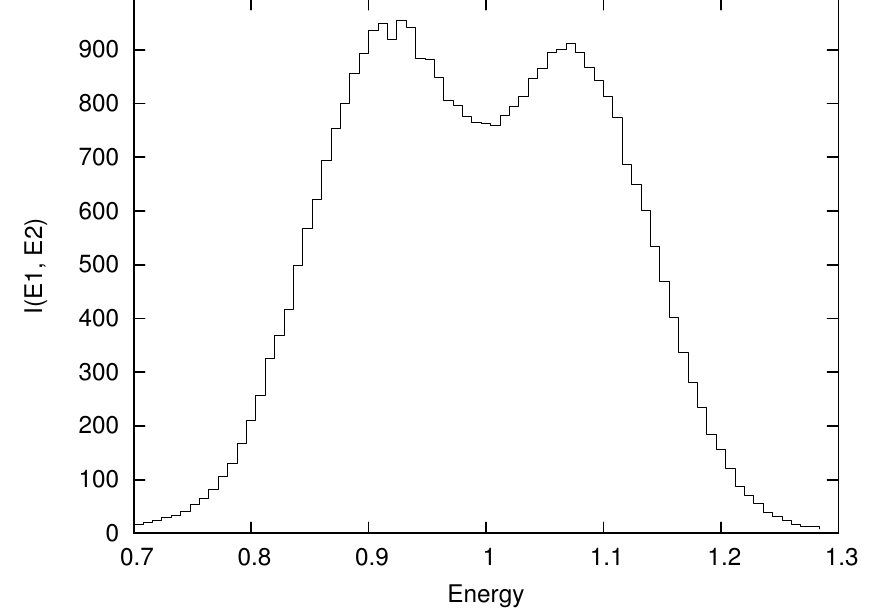}}
\caption{${\cal{I}}(r_{cut} = 34, E_1, E_2)$ for $a = 0$\label{fig:a0inspec200}}
\end{figure}
The shape of the spectrum can also be considered for a non-rotating black hole and for a reasonable value of $r_{cut}$ gives a flux very similar to the Kerr case. Figure.~\ref{fig:a0inspec200} shows the spectrum for a non-rotating black hole with $r_{cut} = 34$ and differs from Figure.~\ref{fig:inspec1-34} by having a slightly larger flux and the peaks closer to $E = 1$. This is due to the smaller Lorentz boosts associated with the Schwarzschild black hole and a faster growing escape fraction. This is found in contrast to the spectrum in Ref.~\cite{Baushev} where the spectrum was found to have a narrow single peak. In considering the incoming dark matter particles to have momenta distributed over the whole range of allowed in falling geodesics the Lorentz shifting of the peak is found to have a splitting effect giving rise to two peaks that accounts for this difference.
\section{Conclusions}
\label{sec:conclusion}
In conclusion we have shown that annihilation of dark matter to massless particles around a black hole can produce a distinctive signal independent of the particle model. The spectrum is found to be centred around the dark matter mass with a splitting into two peaks due to the Lorentz boosting of the centre of mass frame of the colliding particles. The shape of the spectrum retains its distinctive shape when integrated over a typical density profile for the dark matter and can in fact reveal details of the distribution, in particular the radius of the annihilation plateau thought to form around a black hole. The shape of the spectrum is not dependent on the total flux of massless particles from the annihilations which depends on several unknown parameters for a particular source and is therefore a useful tool in discriminating such a signal from astrophysical backgrounds.

In the analysis presented here the final state particles are assumed to be massless, however the final result is not changed for massive particles so long as the mass is small compared to the mass of the colliding particles. This will certainly be true for neutrinos in models where the dark matter mass is of the order of few GeV and above. The numerical calculation can be trivially expanded to include a mass term for the escaping particles and we find little change in the resulting spectrum even for final state particles with a mass half that of the dark matter particle.

In calculating the energy spectrum we assumed that the annihilation cross-section of the dark matter was independent of the centre of mass energy, for a particular dark matter model the energy dependence of the cross section could modify the shape of the spectrum. The cross-section would need to be found for each Monte-Carlo generated collision as the centre of mass energy changes for each pair of colliding particles. We leave this extension for future work. 

In the numerical analysis carried out the collisions were assumed to take place in the equatorial plane and the escape fraction and spectrum were taken to be independent of the initial polar angle $\theta_0$ at which the collision takes place. The results were checked for collisions at $\theta_0 = 0.6$ for $r_{cut} = 34$ and the spectrum retains the same structure with a lower total flux but the same separation between peaks. It would therefore appear that the radial dependence of the energy spectrum dominates over any $\theta_0$ dependence. For small $r$ the spectrum will differ due to the dependence of the radius of the ergosphere on $\theta$ however for reasonable value of $r_{cut}$ the spectrum is dominated by collisions far beyond the ergosphere. 

The results also demonstrate that while large centre of mass energies are possible in collisions close to the horizon the emitted massless particles will in general be largely red shifted to energies around the dark matter mass. High energy massless particles were found in the spectrum with energies far larger than the dark matter mass due to the Penrose process, the non-zero escape fraction for these particles means that collisions around black holes could still be a source of energetic photons or neutrinos with energies several times the dark matter mass, but the flux of these will be small.
\section{Acknowledgements}
We thank Joseph Silk and Stephen West for useful discussions. This work was supported by the Higher Education Funding Council for England and the Science and Technology Facilities Council under the SEPNet Initiative.

\end{document}